\documentclass{iau}
\usepackage{graphicx}

\title[JD 11.~~Radio properties of H$_2$O megamaser Seyfert 2s] 
{A systematic observational study of radio properties of H$_2$O megamaser Seyfert 2s: \\
A Guide for H$_2$O megamaser surveys}

\author[J. S. Zhang, Z. W. Liu, \& C. Henkel]   
{J. S. Zhang$^1$, Z. W. Liu$^1$,
 \and C. Henkel$^{2,3}$}

\affiliation{$^1$Center for Astrophysics, Guangzhou University, \\ Guangzhou 510006, China \\ email: {\tt jszhang@gzhu.edu.cn}
 \\[\affilskip]
$^2$Max-Planck-Institut f{\"u}r Radioastronomie, Auf dem H{\"u}gel 69, \\ D-53121 Bonn, Germany
\\[\affilskip]
$^3$Astron. Dept., King Abdulaziz University, P.O. Box 80203, \\ Jeddah 21589, Saudi Arabia}

\pubyear{2017}
\volume{336}  
\jname{Astrophysical Masers: Unlocking the Mysteries of the Universe}
\editors{A. Tarchi, M.J. Reid \& P. Castangia, eds.}
\begin{document}

\maketitle

\begin{abstract}
Analyzing archival data from different telescopes, H$_2$O megamaser Seyfert 2s appeared to exhibit higher nuclear radio luminosities than non-masing Seyfert 2s (Zhang et al. 2012). This has been confirmed by our follow-up study on multi-band (11, 6, 3.6, 2, 1.3\,cm) radio properties of maser host Seyfert 2s, through systematic Effelsberg observations (Liu et al. 2017). The nuclear radio luminosity was supposed to be a suitable indicator to guide future AGN maser searches. Thus we performed a pilot survey with the Effelsberg telescope on H$_2$O maser emission toward a small sample of radio-bright Seyfert 2 galaxies with relatively higher redshift ($>$0.04). Our pilot survey led to one new megamaser source and one additional possible detection, which reflects our success in selecting H$_2$O megamaser candidates compared to previous observations (higher detection rate, larger distance). Our successful selection technique choosing Seyfert 2s with radio-bright nuclei may provide good guiding for future H$_2$O megamaser surveys. Therefore we are conducting a large systematic survey toward a big Seyfert 2 sample with such radio-bright nuclei. Detections of luminous H$_2$O masers at large distance (z$>$0.04) may hold the great potential to increase our knowledge on the central highly obscured but still very enigmatic regions of active Seyfert galaxies (Zhang et al. 2017).
\end{abstract}

\firstsection 
\section{Introduction}

H$_2$O megamasers are found to be mostly located in heavily obscured nuclear region of Seyfert 2s or LINER galaxies (Braatz et al. 1997, Zhang et al. 2006, Greenhill et al. 2008, Zhang et al. 2010). Due to their extremely high luminosities ($L_{\rm H_{2}O}>10\,L_{\odot}$, assuming isotropic radiation), the ultimate energy source of H$_2$O megamasers is believed to be an AGN, which is supported by all interferometrical studies of megamasers so far carried out (Lo 2005). It was proposed that the nuclear radio continuum may provide ``seed'' photons, which can be amplified by foreground maser clouds to produce strong megamaser emission (E.g., Braatz et al. 1997; Henkel et al. 1998; Herrnstein et al. 1998). In addition, the isotropic luminosity of the nuclear radio continuum is believed to be an indicator of AGN power (e.g., Diamond-Stanic et al. 2009). Therefore, maser emission is expected to be stronger in radio-bright nuclei. The first statistical study on this proposition from Braatz et al. (1997) supported this trend, but with limited statistical significance due to  the small sample (among the 16 known megamaser sources at that time, only nine had the complementary radio data). Now, the situation has greatly improved, so that we can perform a systematic study on the radio properties of H$_2$O megamaser host galaxies.

\section{Analysis and results}

As our first step, the radio continuum data of all published H$_2$O maser galaxies at that time (85 sources) and a complementary Seyfert sample devoid of detected maser emission were investigated and analyzed. Our analysis indicated that maser host Seyfert 2 galaxies have higher nuclear radio continuum luminosities (at 6\,cm and 20\,cm), exceeding those of the comparison Seyfert 2 sample by factors of order 5 (Figure\,1). This supports the previous proposition that the nuclear H$_2$O megamaser emission is correlated with the nuclear radio emission and the nuclear radio luminosity may be a suitable indicator to guide future AGN maser searches (Zhang et al. 2012).

\begin{figure}[b]
\centering \mbox{
\includegraphics[bb=320 570 0 400, width=9.0cm, angle=-180]{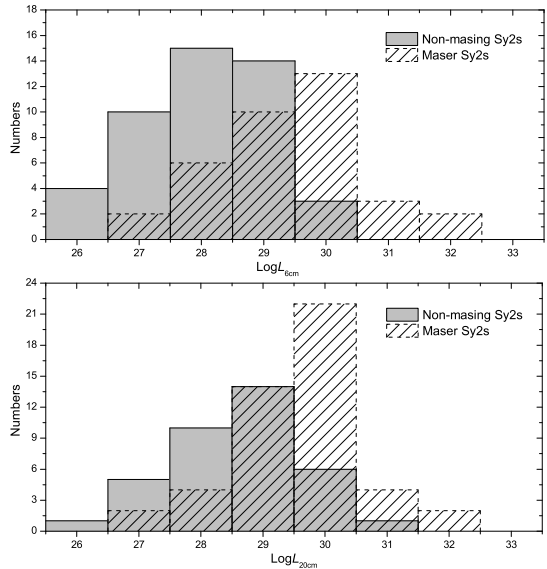}}
\vspace*{3.5 cm}
 \caption{Radio luminosity distributions for maser and non-masing Sy2s (in erg\,s$^{-1}$\,Hz$^{-1}$).}
\end{figure}

However, the uncertainties of this initial analysis were still quite large. Measured data (at 20\,cm and 6\,cm) were mostly taken from different telescopes for both maser sources and non-masing sources. Even if data were taken from the same telescope, measurements were normally performed at different epochs. And there are only a few data at other radio bands for both samples (e.g., 3.6\,cm, 2.0\,cm). A more complete radio dataset was urgently needed. While an interferometric study would be a better choice, systematic studies of the lower resolution radio continuum with single-dish telescopes are still worthwhile, because they are not affected by missing flux.

As the second step, we therefore performed systematic Effelsberg multi-band radio continuum observations (11\,cm, 6.0\,cm, 3.6\,cm, 2.0\,cm and 1.3\,cm) within a tiny time span in January 2014 toward the H$_2$O megamaser Seyfert 2s and the control Seyfert 2 sample without detected maser emission. Our analysis shows that a difference in radio luminosity (at all bands) is statistically significant, i.e. that the maser Seyfert 2 galaxies tend to have higher radio luminosities by a factor of 2 to 3 than the non-masing ones, commonly reaching values above a critical threshold of 10$^{29}$\,erg\,s$^{-1}$\,Hz$^{-1}$ (Figure 2). The difference between maser and non-masing Seyfert 2s is supported by observations in each wavelength band (bottom panel in Figure 2), i.e., the mean flux density of maser Seyfert 2s is larger than those of the non-masing ones at each band, roughly by a factor of 2. In addition, the black hole mass and the accretion rate were estimated for our maser Seyfert 2s and non-masing Seyfert 2s, assuming the radio luminosity as an isotropic tracer of AGN power. It shows that the accretion rates of maser Seyfert 2s are nearly one order magnitude larger than those of non-masing Seyfert 2s (Liu et al. 2017). This may provide a possible connection between H$_2$O megamaser formation and AGN activity, as well as suitable constraints on future megamaser surveys.

\begin{figure}
\mbox{
\includegraphics[bb=320 670 0 400, width=12.0cm, angle=-180]{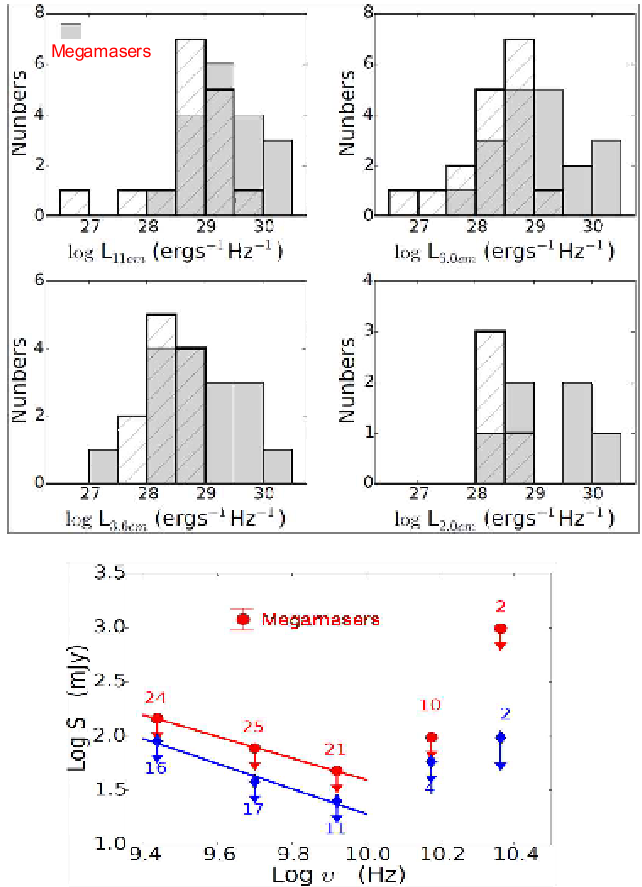}}
\vspace*{2.5 cm}
 \caption{Upper panels: Comparisons of radio luminosities for megamaser and non-masing Sy2s. Bottom: The SED for maser Sy 2s and non-masing Sy2s. Trend lines at low frequencies (i.e. 11 cm, 6.0 cm and 3.6 cm) are also presented for both samples, from Liu et al. (2017).}
\end{figure}

\begin{figure}
\centering 
\vspace{0.5cm}

 \includegraphics[width=9cm]{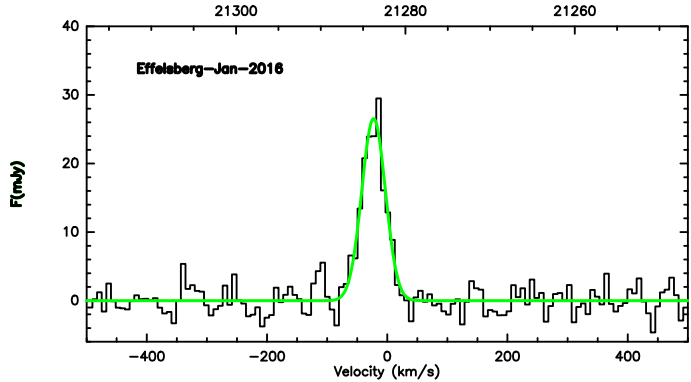}
 \caption{The smoothed average spectrum with the fitting line of the new megamaser source SDSS\,102802.9+104630.4, with a redshift of 0.044776, taken from Zhang et al. (2017).}
\end{figure}

From previous analysis of archival radio data (Zhang et al. 2012) and systematic observations with the Effelsberg telescope (Liu et al. 2017), we found strong evidence for a scenario where H$_2$O megamasers locate in radio-bright Seyfert 2s. Nuclear radio luminosity is therefore supposedly a suitable indicator to guide future AGN maser searches. To test this, we conducted, as a third step, a pilot survey toward a small sample of Seyfert 2s with radio-bright nuclei, using the Effelsberg-100 m telescope. Our 18 targets were selected from a large SDSS-DR7 Seyfert 2 sample containing 4035 sources (0.04$<$z$<$0.1, Coldwell et al. 2013), with a 20\,cm luminosity threshold of 10$^{29}$\,erg\,s$^{-1}$\,Hz$^{-1}$ (Liu et al. 2017). Even though our survey was, due to the large distance of the targets, extremely shallow, it led to one new strong megamaser source (L$_{\rm H2O}$$>$1000\,L$_{\odot}$, Figure 3) and one additional tentative detection with an (isotropic) luminosity of several 100\,L$\,_{\odot}$, which reflects our success in selecting H$_2$O megamaser candidates compared to previous observations (Zhang et al. 2017). While our detection rate of 5.5\% or 11\% (the latter including a tentative detection) appears to be higher than in most other surveys (the uncertainty lies in the limited number of studied sources), the distance to our targets is also larger, thus making the success of this pilot survey particularly noteworthy. Our successful selection technique may provide good guiding for future H$_2$O megamaser surveys, i.e., Seyfert 2s with radio-bright nuclei.

Thus we will continue our H$_2$O megamaser survey toward that large Seyfert 2 sample, with our radio luminosity criterium of log\,$L$$>$29\,erg\,s$^{-1}$\,Hz$^{-1}$. More H$_2$O megamaser detections at a large distance (z$>$0.04) can be expected from such a large systematic survey of radio-bright Seyfert 2 galaxies. While presumably leading to a better determination of the upper (isotropic) luminosity limit of such masers, these will shed new light onto the central highly obscured but very enigmatic nuclear regions of active Seyfert galaxies.

\begin{acknowledgements}

This work is supported by the Natural Science Foundation of China (No. 11473007, 11590782). Thank the Effelsberg staff much for their kind help during our observations.

\end{acknowledgements}
\end{document}